%% file: main.tex
\documentclass[runningheads]{llncs}

\usepackage[T1]{fontenc}

\usepackage{graphicx}
\usepackage{amsmath}

\usepackage{latexsym}
\usepackage{enumitem}
\usepackage{microtype}
\usepackage{inconsolata}
\usepackage{graphicx}
\usepackage{algorithm}
\usepackage{algpseudocode}
\usepackage{hyperref}

\usepackage{xcolor}

\begin{document}

\title{Misconception Acquisition Dynamics \\ in Large Language Models
}
\titlerunning{Misconception Acquisition Dynamics in LLMs}



\author{Naiming Liu$^{*}$\inst{1} \and
Xinghe Chen$^{*}$\inst{1} \and
Richard Baraniuk\inst{1} \and \\
Mrinmaya Sachan\inst{2} \and
Shashank Sonkar\inst{3}
}

\authorrunning{N. Liu et al.}

\institute{Rice University \and
ETH Zürich \and
University of Central Florida \\ 
\email{nl35@rice.edu, shashank.sonkar@ucf.edu}
}

\maketitle              

\begingroup
\renewcommand\thefootnote{*}
\footnotetext{Equal Contribution.}
\endgroup
\setcounter{footnote}{0}

\begin{abstract}
\input{sections/abstract}

\keywords{Large Language Models \and Instruction Tuning \and Student Misconceptions \and Learner Modeling}
\end{abstract}


\section{Introduction}
\input{sections/intro}

\section{Related Work}
\input{sections/related_work}

\input{tables/pt_w_arch}
\section{MalAlgoLib: Algebraic Problem-Solving with Misconceptions}
\input{sections/library}

\section{Formalizing Misconception Acquisition}
\input{sections/method}

\section{Experiments}
\input{sections/experiments}

\section{Conclusion}
\input{sections/conclusion}

\section*{Acknowledgments}
This work was supported by NSF SafeInsights and cooperative agreement 2153481.
%
%
\bibliographystyle{splncs04}
\bibliography{custom}

\end{document}

%% file: sections/abstract.tex
Effective educational AI depends on modeling student misconceptions.
Such models enable realistic learner simulation and diagnostic, adaptive tutoring.
However, instruction-tuning large language models (LLMs) on student responses containing misconception errors can degrade reasoning abilities, creating a tension between faithful misconception modeling and preserving correct reasoning in other contexts.
To support both learner simulation and tutoring, we study two misconception-aware models: the Novice Student Misconception Model, trained to acquire a single misconception for simulating an individual student, and the Expert Tutor Misconception Model, trained on multiple misconceptions to capture the error patterns a tutor encounters across students.
To study the misconception acquisition dynamics of both models, we develop MalAlgoLib, a library that generates algebra problems with correct solution traces and misconception-specific erroneous traces.
Our experiments across three LLMs reveal that the student and the tutor model exhibit fundamentally different misconception acquisition dynamics.
For the student model, a single misconception is not learned as a context-specific behavior. Models overapply it across problems, degrading correct-solving accuracy unless training includes correct examples to enforce boundaries.
In contrast, the tutor model can learn multiple misconceptions jointly without sacrificing correct-solving accuracy. 
Critically, intermediate reasoning steps are the bottleneck. With final-answer supervision alone, models cannot learn where error enters the solution, so neither the student model nor the tutor model acquires misconceptions regardless of data size.
Together, these results, enabled by MalAlgoLib, provide an interpretable account of misconception acquisition under instruction tuning and guidance for training misconception-aware LLMs while preserving correct reasoning.

%% file: sections/intro.tex
Large language models (LLMs) are transforming educational technology along two complementary directions. On one hand, LLMs enable sophisticated student modeling: simulating learner behavior for intelligent tutoring systems~\cite{liu2024personality,benedetto2024llmstudent,liu2023novice} and generating realistic student responses for teacher training~\cite{piech2023gpteach,leegenerative,chen2026}. On the other hand, LLMs can support teachers and tutors directly: providing personalized feedback~\cite{pal2024autotutor,class}, identifying student difficulties, and adapting instruction to individual needs~\cite{schmucker2023ruffle,pedalign}. Both directions share a critical requirement: understanding student \emph{misconceptions}, the systematic errors that reflect underlying conceptual misunderstandings. 

Understanding these error patterns is essential for effective instruction. Research shows that teaching that directly targets misconceptions produces stronger learning gains than simply repeating correct procedures~\cite{brown1978diagnostic}. This works because these patterns are not random but persistent and predictable, arise from flawed reasoning strategies~\cite{malrules,sleeman}. For example, when solving $ax = b(cx + d)$, students commonly distribute $b$ to only the first term, yielding $ax = bcx + d$ instead of the correct $ax = bcx + bd$. This motivates a fundamental question: \emph{can we instruction-tune LLMs to acquire misconceptions, and what are the training dynamics of this acquisition process?}

Recent evidence suggests this is challenging. MalAlgoQA benchmark~\cite{malalgoqa} revealed that LLMs struggle to identify ``malgorithms'', the flawed reasoning steps that lead to incorrect answers. The EEDI challenge~\cite{king2024eedi} highlighted the difficulty of mapping student misconceptions at scale. 
Most concerning, the student data paradox~\cite{sonkar2024paradox} demonstrated that training LLMs on student dialogue data actually \emph{degrades} their factual knowledge and reasoning abilities. 

These findings show that naively training LLMs on student data is insufficient and can be counterproductive. Yet misconception-aware models are needed in practice, in at least two distinct settings. The first is student modeling, where the goal is to simulate an individual learner exhibiting a specific misconception for tasks such as tutoring-system evaluation and teacher training. The second is tutor modeling, where the goal is to capture the range of misconceptions encountered across many students, providing a computational analogue of teachers’ \emph{Knowledge of Student Misconceptions} (KOSM). This analogue is pedagogically meaningful: teachers' KOSM is linked to improved learning outcomes beyond subject-matter knowledge alone~\cite{sadler2020kosm}. We formalize these as the \textbf{Novice Student Misconception Model}, which acquires a single misconception, and the \textbf{Expert Tutor Misconception Model}, which acquires multiple misconceptions simultaneously.

To systematically study what LLMs need in order to acquire misconceptions, we require controlled data that is both cognitively faithful and available at scale, since LLM instruction tuning is data-hungry. 
We therefore build on a long line of cognitive-science work that treats student errors as systematic products of incorrect procedures.
In this ``buggy-rule'' or ``mal-rule'' view, a misconception is a reusable operator that modifies an otherwise correct algebraic transformation at a specific step (for example, distributing across only one term mentioned earlier), producing predictable wrong intermediate states~\cite{malrules,sleeman,sleeman2}.
We developed MalAlgoLib which operationalizes this idea by representing equation solving as a graph of problem types and encoding each misconception as an alternative step that yields its corresponding erroneous trace. 
MalAlgoLib covers 16 linear-equation problem types and 20 misconceptions drawn from established taxonomies~\cite{malrules,sleeman,sleeman2} enabling scalable generation of paired correct and misconception-driven solution traces for studying misconception acquision dynamics.

For our experiments, we instruction-tuned three LLMs (Llama-8B, Phi-4-4B, and Qwen-4B~\cite{llama3,phi4,qwen3}). We find that the student model and the tutor model exhibit fundamentally different training dynamics. For the student model, models fail to localize a learned misconception to relevant contexts and instead overgeneralize it across problems, which degrades correct-solving accuracy. To obtain faithful student models, correct examples must be explicitly mixed with misconception data: even a small fraction of correct examples (as low as 25\%) allows models to reproduce the target misconception on the intended problem types while preserving correct reasoning on problems where the misconception should not apply.

Tutor model training dynamics differ markedly. When trained on ten misconceptions simultaneously, models learn to replicate each misconception while their correct problem-solving ability remains stable. Unlike the student model, the tutor model does not require explicit mixing of correct examples. However, sample efficiency poses a practical challenge: classroom-scale data (5--80 samples per misconception) proves insufficient for reliable acquisition. As Payne and Squibb~\cite{malrules} documented, misconception frequencies are heavily skewed even within a single school, meaning rare misconceptions may not occur often enough to provide sufficient training examples. Real-world tutor model training would therefore likely require aggregating data across multiple educational institutions.

Together, these observations situate MalAlgoLib as a controlled testbed for studying misconception acquisition in the high-data regime that the tutor model requires. Critically, we find that \textbf{step-level supervision is essential for both models}. Without intermediate solution steps, neither the student model nor the tutor model can acquire misconceptions regardless of training data size. Training on final incorrect answers alone does not show where the misconception enters the solution, so models fail to acquire the misconception. MalAlgoLib's synthetic step-by-step traces enable us to study these dynamics in a controlled setting. However, real-world deployment faces a significant challenge: student data collected from assessments typically contains only final answers, and obtaining step-by-step reasoning at scale requires privacy-preserving infrastructure. Emerging approaches such as secure data enclaves~\cite{safeinsights} and federated learning offer promising directions for collecting the rich, step-level student data that our findings suggest is essential for misconception-aware educational AI.

\input{tables/misc_2_pt_map}

In summary, this work makes three contributions. First, we formalize the task of \emph{misconception acquisition} and define two complementary models, the \textbf{Novice Student Misconception Model} and the \textbf{Expert Tutor Misconception Model}, that address distinct educational needs. Second, through comprehensive experiments across three LLMs, we characterize the acquisition dynamics of both models and provide practical recommendations: mix correct data for the student model, aggregate multi-source data for the tutor model, and always include step-level supervision. Third, we introduce \textbf{MalAlgoLib}, a graph-based library for generating cognitively-grounded algebra datasets with step-by-step solution traces, which enabled the controlled experiments underlying these findings. 
Code is available \href{https://github.com/sonkar-lab/MalAlgoLib}{here}.

%% file: tables/misc_2_pt_map.tex
\begin{table*}[t]
\centering
\caption{Algebraic misconceptions modeled in MalAlgoLib. MalAlgoLib implements problem types and misconceptions from Payne and Squibb's foundational mal-rules taxonomy~\cite{malrules} for controlled misconception data generation.}
\resizebox{\textwidth}{!}{
\begin{tabular}{|l|l|l|l|}
\hline
\textbf{Misconception} & \textbf{Expression} & \textbf{Applicable Types} & \textbf{Description} \\
\hline
M1 & $A(part) \to A + (part)$ & T8, T9, T10, T12 & Treating distribution as addition \\
M2\_S3 & $A(Bx \pm C) \to ABx \pm C$ & T9, T12 & Ignoring distribution \\
M3 & $A \pm B(part) \to (A \pm B)(part)$ & T10, T12 & Misapplying parentheses \\
M4 & $A(B*C) \to A*B*A*C$ & T8 & Incorrectly distributing multiplication \\
M5 & $A(Bx \pm C) \to A(A*Bx \pm A*C)$ & T9, T12 & Over-distribution \\
M6 & $-A(Bx - C) \to -A*Bx - A*C$ & T9, T12 & Incorrect sign distribution \\
M8 & $A(Bx \pm C) \to Bx \pm A*C$ & T9, T12 & Incorrect distribution on x term \\
M11 & $Ax \pm B = Cx \pm D \to Ax + Cx = B + D$ & T14 & Incorrectly combining terms \\
M12/S15 & $Ax \pm B = (A \pm B)x$ & T5, T6, T7, T9, T12 & Incorrectly factoring x \\
M13 & $Ax \pm B = (A \pm B)$ & T5, T6, T7, T9, T12 & Incorrectly factoring x \\
M14 & $part1 + part2 \to part1 - part2$ & T2, T4 &  Incorrect swap of addition \& subtraction \\
M15 & $part1 - part2 \to part1 + part2$ & T2, T4 &  Incorrect swap of addition \& subtraction \\
M16 & $part1 * part2 \to part1 + part2$ & T3, T10 & Treating multiplication as addition \\
M17 & $A + B \to B - A$ & T2, T4 & Incorrect swap of addition \& subtraction \\
M18 & $A - B \to B - A$ & T2, T4 & Incorrect swap of addition \& subtraction \\
M19 & $Ax = B \to x = A + B$ & All Types & Treat division as addition \\
M20\_S20 & $Ax = B \to x = B$ & All Types & Divide only on one side \\
M21 & $Ax = B \to x = A - B$ & All Types & Treat division as subtraction \\
M22\_S1 & $Ax = B \to x = A/B$ & All Types & Incorrect numerator and denominator \\
\hline
\end{tabular}
}
\label{tab:misc}
\end{table*}

%% file: sections/related_work.tex
\label{sec:related}

LLMs are increasingly deployed in educational settings. Recent work has explored using LLMs for student simulation, where models generate responses that mimic learners of varying skill levels~\cite{benedetto2024llmstudent,piech2023gpteach}. Other efforts have focused on intelligent tutoring, developing LLM-based systems that provide pedagogically-grounded feedback and adaptive instruction~\cite{pal2024autotutor,schmucker2023ruffle}. While these approaches have shown promise, they typically focus on modeling correct behavior. The question of whether LLMs can accurately acquire and replicate specific student misconceptions has received less attention.

The study of student misconceptions has a rich history in cognitive science and mathematics education. Foundational work on problem-solving established frameworks for understanding human reasoning~\cite{newell1972human,anderson1993rules}. Brown and Burton's BUGGY system~\cite{brown1978diagnostic} pioneered computational diagnosis of procedural errors in arithmetic, demonstrating that student mistakes follow systematic patterns. For algebra, Sleeman~\cite{sleeman,sleeman2} and Matz~\cite{matz1982towards} documented common error types and their cognitive origins, while Payne and Squibb~\cite{malrules} formalized these as ``mal-rules,'' incorrect but consistently applied procedures. The Cognitive Tutor~\cite{ritter2007cognitive} successfully translated these insights into deployed educational software. However, these classical approaches rely on hand-crafted rules rather than learning misconceptions from data.

Student modeling research has developed sophisticated methods for tracking learner knowledge over time. Knowledge tracing~\cite{corbett1994knowledge} models the probability that a student has mastered a skill based on their response history. Bayesian networks~\cite{conati2002using} extend this framework to handle uncertainty and complex skill dependencies. Comprehensive surveys~\cite{baker_test} document the evolution of these methods. While effective for tracking mastery, these approaches typically do not model the specific nature of student errors; they track whether students answer correctly, not why they answer incorrectly. Our work takes a complementary approach by explicitly modeling how misconceptions are acquired and applied.

Recent work has begun to examine LLMs through an educational lens, revealing both opportunities and challenges. 
The MalAlgoQA benchmark~\cite{malalgoqa} evaluated LLMs' ability to identify flawed reasoning in algebraic problem-solving, revealing significant limitations. Most relevant to our work, Sonkar et al.~\cite{sonkar2024paradox} documented the Student Data Paradox: training LLMs on student dialogue data degrades their factual knowledge and reasoning abilities. This finding motivates careful study of how to train LLMs on misconception data without compromising core capabilities. Educational research has also shown that teachers with Knowledge of Student Misconceptions produce better learning outcomes~\cite{sadler2020kosm}, suggesting value in models that can recognize diverse error patterns. Our work addresses these challenges by formalizing misconception acquisition as a learning problem, defining complementary models for student and tutor perspectives, and systematically studying their instruction-tuning dynamics.

%% file: tables/pt_w_arch.tex
\begin{figure*}[t]
\begin{minipage}{0.3\textwidth}
\centering
\resizebox{\linewidth}{!}{
\begin{tabular}{|l|l|}
\hline
\textbf{Type} & \textbf{Expression} \\
\hline
T1 & $Ax = B$ \\
T2 & $Ax = B + C$ \\
T3 & $Ax = B * C$ \\
T4 & $Ax + Bx = C$ \\
T5 & $Ax + B = C$ \\
T6 & $A + Bx = C$ \\
T7 & $Ax = Bx + C$ \\
T8 & $Ax = B(C*D)$ \\
T9 & $Ax = B(Cx + D)$ \\
T10 & $Ax = B + C * D$ \\
T11 & $A + Bx + Cx = D$ \\
T12 & $Ax = B + C(Dx + E)$ \\
T14 & $Ax + B = Cx + D$ \\
T15 & $Ax + Bx = C + D$ \\
T16 & $Ax = Bx + C + D$ \\
\hline
\end{tabular}
}
\label{tab:problem_types}
\end{minipage}%
\begin{minipage}{0.7\textwidth}
\centering
\includegraphics[width=\linewidth]{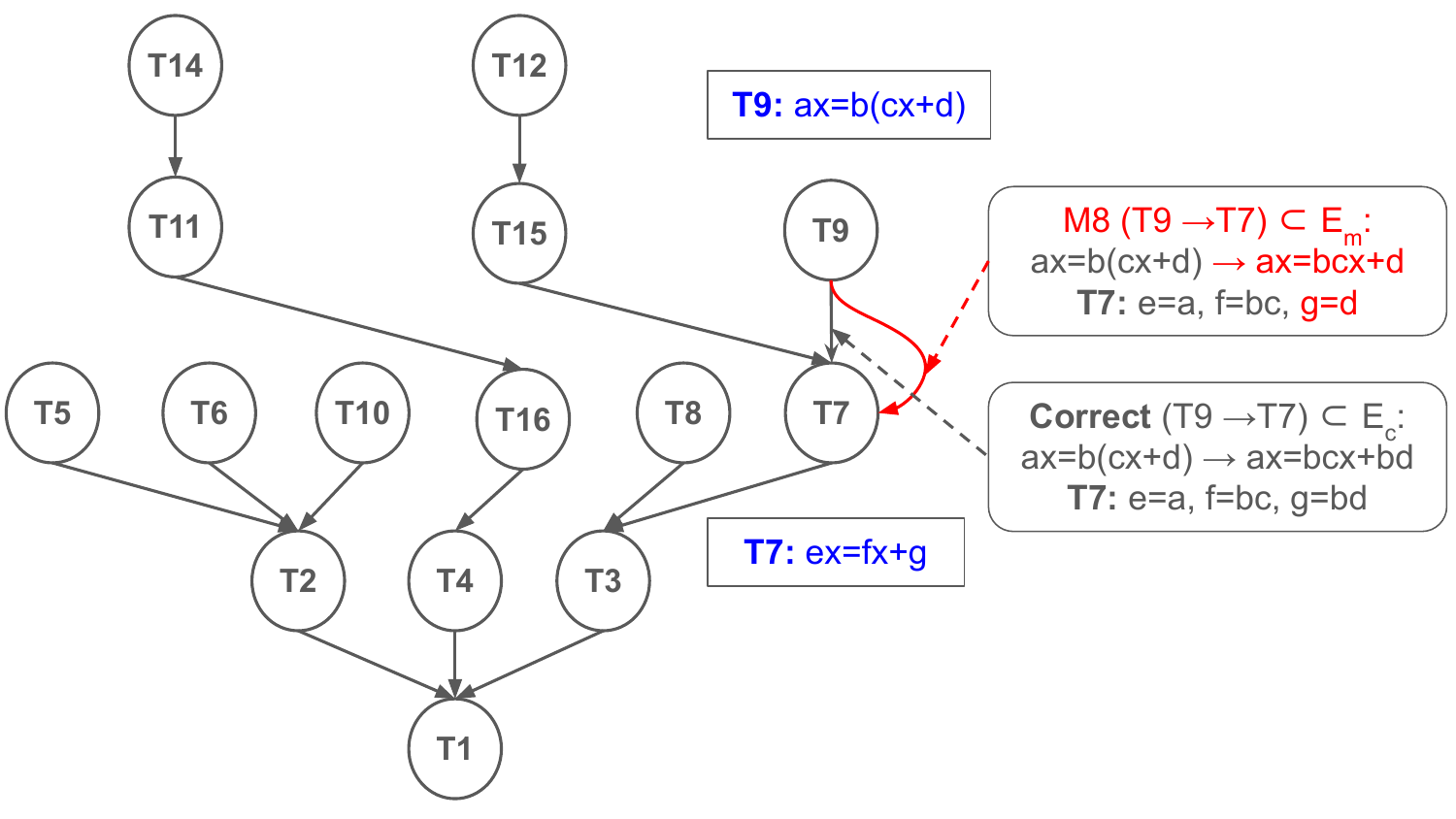}
\end{minipage}
\caption{\textbf{Left:} Problem types in MalAlgoLib. \textbf{Right:} Graph-based model where nodes are problem types. Each gray edge represent a reduction, similar to students' steps, from complex type to a more simple type eventually reaching the base type T1. Each red edge (only M8 shown) represents a step reduction conducted under a misconception. Such reduction structure reflected our design principle of code reuseability, easy expansion, and closely mirrors student's cognitive process of equation solving.}
\vspace{-10pt}
\label{fig:tree}
\end{figure*}

%% file: sections/library.tex
\label{sec:MalAlgoLib}
Studying the dynamics of instruction-tuning LLMs to acquire misconceptions poses two data challenges. First, real student assessments typically capture only final answers, not the intermediate reasoning steps our experiments will show are essential. Second, misconception frequencies are heavily skewed in real classrooms~\cite{malrules}, making it difficult to collect balanced training data. MalAlgoLib addresses these challenges by generating synthetic algebraic problem-solving data with step-by-step solution traces, enabling controlled experiments that systematically vary training data composition. Its architecture is built on three design principles: extensibility across equation types, modularity in misconception modeling, and faithful representation of cognitive error patterns.

\subsection{Problem Type Hierarchy and Reduction Process}
\label{sec:core_intro}

MalAlgoLib (Mal-Algorithm Library) is designed to handle linear equations with one variable, including more complex forms with parentheses and multiple terms.
At its core is a type-based problem representation implemented as a directed acyclic graph $G = (V, E)$. The set $V = \{T_1, T_2, \ldots, T_{16}\}$ represents a hierarchy of problem types, where each $T_i \in V$ corresponds to a specific equation structure. Utilizing such graph-like structure, we enforces a strict single-step reduction policy, closely mirroring the student cognitive process of step-by-step problem-solving. Each type class implements a \texttt{reduce()} method, represents a specific algebraic operation, that applies a single algebraic transformation, transitioning the equation to a less complex type. Addition of new problem type would be simply utilize the template and define \texttt{reduce()} method to link to an existing problem type. 

\subsection{Decoupled Misconception Modeling}

The decoupled misconception modeling, where misconceptions are represented by alternative transitions between problem types (Figure~\ref{fig:tree}), implement misconceptions fitting our three design principle. This approach allows for flexible and extensible modeling of student errors within the graph-based framework. It enables cross-type applicability, where a single misconception can be applied across multiple equation types, and combinatorial error generation, where the graph structure allows for the natural combination of multiple misconceptions along a solution path. Additionally, this design facilitates easy extensibility, as new misconceptions can be added by introducing new alternative transitions without modifying existing problem-type implementations, supporting ongoing research and refinement of the error model.

These three principles enable MalAlgoLib to generate controlled training data for studying misconception acquisition. The system produces step-by-step solution traces for both correct and misconception-based problem-solving, with precise control over data composition. This enables the experiments in Section~\ref{sec:exps}, where we systematically vary training data size, misconception combinations, and the presence of intermediate steps to characterize the instruction-tuning dynamics of the Novice Student Misconception Model and the Expert Tutor Misconception Model.

%% file: sections/method.tex
\label{sec:csa}

We study the dynamics of instruction-tuning LLMs to acquire misconceptions from two complementary perspectives. First, the \textbf{student perspective}: can we train LLMs to simulate individual students who hold specific misconceptions? Second, the \textbf{tutor perspective}: can we train LLMs to understand the full spectrum of misconceptions that tutors encounter across their students? These perspectives lead to fundamentally different model requirements and, as we will show, exhibit markedly different learning dynamics. We formalize both perspectives as the Novice Student Misconception Model and the Expert Tutor Misconception Model, define evaluation metrics, and investigate how training data composition affects each model's ability to acquire misconceptions while maintaining correct problem-solving capabilities.

\subsection{Evaluation Metrics}

We define the following metrics:

\textbf{Misconception Accuracy (MA)}: Assesses a model's ability to accurately replicate a specific misconception for applicable problem types.
\begin{equation*}
\text{MA}(A, M) = \frac{1}{|\alpha(M)|} \sum_{T_i \in \alpha(M)} \text{MA}(A, M, T_i),
\end{equation*}
where $\text{MA}(A, M, T_i)$ is the accuracy of model $A$ in applying misconception $M$ to problem type $T_i$, $P$ is set of problem types, $\alpha(M) = \{T_i \in P: M \text{ is applicable to } T_i\}$.

\textbf{Correct Accuracy (Applicable) (CA$_A$)}:
Assesses a model's ability to solve problems correctly for problem types where a given misconception is applicable.
\begin{equation*}
\text{CA}_A(A, M) = \frac{1}{|\alpha(M)|} \sum_{T_i \in \alpha(M)} \text{CA}(A, T_i).
\end{equation*}

\textbf{Correct Accuracy (Non-Applicable) (CA$_{NA}$)}:
Assesses a model's performance on problem types not applicable to a given misconception.
\begin{equation*}
\text{CA}_{NA}(A, M) = \frac{1}{|P \setminus \alpha(M)|} \sum_{T_i \notin \alpha(M)} \text{CA}(A, T_i).
\end{equation*}

\textbf{Overall Correct Accuracy (OCA)}:
Assesses a model's overall ability to solve problems correctly across all problem types.
\begin{equation*}
\text{OCA}(A) = \frac{1}{|P|} \sum_{T_i \in P} \text{CA}(A, T_i).
\end{equation*}

\subsection{Model Definitions}

We introduce two complementary models that address different educational needs. The \textbf{Novice Student Misconception Model} simulates how an individual student with a specific misconception approaches algebraic problems, useful for understanding how misconceptions manifest in problem-solving and for generating realistic student responses. The \textbf{Expert Tutor Misconception Model} captures the full spectrum of misconceptions that a tutor might encounter across their students, enabling tutors to recognize and address diverse error patterns at scale. Hereafter, we refer to these as the \emph{student model} and the \emph{tutor model}, respectively.

\subsubsection{Novice Student Misconception Model}

The student model focuses on acquiring a single misconception while preserving correct reasoning on unaffected problem types. This mirrors how an individual student might consistently apply one flawed rule (e.g., distributing incorrectly) while solving other problem types correctly.

\begin{definition}[Novice Student Misconception Model]
A model $A$ successfully acquires misconception $M$ if it satisfies two properties:

\noindent\textbf{Property 1 (Misconception Replication):} For problem types where $M$ applies, the model accurately reproduces the misconception:
\begin{equation*}
\text{MA}(A, M) \geq \theta_m
\end{equation*}

\noindent\textbf{Property 2 (Correct Reasoning Preservation):} For problem types where $M$ does not apply, the model solves problems correctly:
\begin{equation*}
\text{CA}_{NA}(A, M) \geq \theta_c
\end{equation*}
\end{definition}

\subsubsection{Expert Tutor Misconception Model}

The tutor model addresses a more challenging scenario: acquiring multiple misconceptions simultaneously while maintaining overall problem-solving accuracy. Unlike the student model, which only needs to preserve accuracy on non-applicable problems, the tutor model must solve \emph{all} problem types correctly when prompted to do so---even those affected by misconceptions in its training set. This reflects how an expert tutor understands various student errors yet can demonstrate correct solutions when needed.

\begin{definition}[Expert Tutor Misconception Model]
A model $T$ successfully acquires a misconception set $M_{set}$ if it satisfies two properties:

\noindent\textbf{Property 1 (Multi-Misconception Replication):} For every misconception $M_i \in M_{set}$, the model accurately reproduces it on applicable problem types:
\begin{equation*}
\forall M_i \in M_{set}: \text{MA}(T, M_i) \geq \theta_m
\end{equation*}

\noindent\textbf{Property 2 (Overall Correct Accuracy):} Across all problem types, the model solves problems correctly when asked:
\begin{equation*}
\text{OCA}(T) \geq \theta_c
\end{equation*}
\end{definition}

\noindent We set $\theta_m = \theta_c = 90\%$ in all experiments.

%% file: sections/experiments.tex
\label{sec:exps}

We now investigate the misconception acquisition dynamics of both the Novice Student Misconception Model (student model) and the Expert Tutor Misconception Model (tutor model). Our central finding is that these two model types exhibit \emph{fundamentally different} instruction-tuning dynamics: student model training degrades correct-solving accuracy, requiring explicit mixing of correct examples, while tutor model training does not, and in fact \emph{improves} overall accuracy. We also demonstrate that step-level supervision is essential for both model types; without it, neither can acquire misconceptions regardless of training data size.

\subsection{Experimental Setup}

\paragraph{Base Models.} We fine-tuned three base models: \texttt{Llama-3.1-8B-Instruct}, \texttt{Phi-4-4B-Mini}, and \texttt{Qwen-3-4B-Instruct}. Each was first trained on 2000 correct examples per problem type to achieve OCA $\geq 90\%$, establishing a foundation of correct problem-solving before introducing misconceptions.

\paragraph{Novice Student Model Training.} For each misconception $M_j$, we trained models on varying amounts of misconception examples $n_m \in \{100, 200, 400, 800, 1600, 3200\}$, measuring both MA and CA$_{NA}$ to assess the acquisition-accuracy trade-off.

\paragraph{Expert Tutor Model Training.} We selected ten representative misconceptions and trained models on $n_m \in \{5, 10, 20, 40, 80\}$ (small-sample regime) and $n_m \in \{100, 200, 400, 800\}$ (standard regime), measuring MA for each misconception and OCA across all problem types.

\paragraph{No-Steps Ablation.} To assess the role of step-level supervision, we repeated experiments with training data containing only final answers (no intermediate steps).

\paragraph{Hyperparameters.} Learning rate $2 \times 10^{-5}$ with AdamW~\cite{loshchilov2019decoupled}, batch size 16, cosine scheduler, weight decay 0.05, warmup ratio 0.1, and 3 epochs. We evaluate on 500 test samples per problem type.

\subsection{Student Model Dynamics: Misconception-Accuracy Trade-off}

\input{figures/misc_acc}
\input{figures/mix_combined}

We first examine student model training dynamics, where a clear trade-off emerges between misconception acquisition and correct-solving accuracy.

Figure~\ref{fig:misconception_vs_correct_accuracy_4plots} provides a comprehensive view of these dynamics across all misconception types. Our analysis of Misconception Accuracy (MA) revealed substantial variation in the number of examples required for different misconceptions to be reliably acquired. While the base models required 2000 correct examples per problem type to reach 90\% OCA, many misconceptions were learned with far fewer samples, whereas others required significantly more data.

For many misconceptions, we observed a characteristic S-shaped learning curve in MA as shown in Figure~\ref{fig:misconception_vs_correct_accuracy_4plots}. As the number of misconception training samples increased, MA rose steadily; however, this improvement was accompanied by a dramatic decline in OCA and $\text{CA}_{NA}$. This inverse relationship is surprising: one might expect models to compartmentalize misconceptions, applying them only to relevant problem types while preserving correct reasoning elsewhere. Instead, naive training causes models to overgeneralize erroneous reasoning patterns, degrading their ability to solve even problems where the misconception should not apply.

These results demonstrate that directly training on a single misconception is typically insufficient to satisfy both Property~1 and Property~2 simultaneously. Instead, strong misconception acquisition frequently comes at the cost of correct reasoning on both applicable and non-applicable problem types. To address this challenge, we trained the student models using $A(M_j, n_m, n_c)$ for misconception $M_j$. For each case, $n_m$ was fixed at the minimum number of misconception examples required to achieve over 90\% MA in the previous experiments. We then introduced additional correct examples, varying the ratio such that $n_c / n_m \in \{0.25, 0.5, 1.0\}$.

Figure~\ref{fig:mix_combiined} shows that, for most misconceptions, balanced training led to substantial recovery in $\text{CA}_{NA}$. Importantly, the introduction of correct examples did not impair the model’s ability to replicate misconceptions when applicable. For example, in the case of the distributional property error (M8), MA remained consistently high, increasing from 90.2\% at 400 samples to 94.5\% at 3200 samples. At the same time, $\text{CA}_{NA}$ remains high, satisfying Property~2. These results demonstrate that carefully mixing correct and misconception examples enables the construction of student models that satisfy both required properties.

\input{figures/etm_combined}
\input{figures/no_steps}

\subsection{Tutor Model Dynamics: No Accuracy Trade-off}

We now turn to the tutor model, where the learning dynamics differ markedly. The tutor model must acquire multiple misconceptions simultaneously, a capability essential for tutors who must recognize diverse error patterns across their students. Unlike the student model, which focuses on a single misconception, the tutor model must satisfy a more demanding criterion: maintaining high OCA across \emph{all} problem types (not just non-applicable ones) while accurately replicating each misconception in its training set.

We selected a representative set of ten misconceptions:
\[
M_{set} = \{M_{1}, M_{2}, M_{6}, M_{8}, M_{11}, M_{12}, M_{19}, M_{20}, M_{21}, M_{22}\},
\]
and trained models on varying amounts of data per misconception:
\[
n_m \in \{5, 10, 20, 40, 80\} \text{ (small-sample regime)}
\]
\[
n_m \in \{100, 200, 400, 800\} \text{ (standard regime)}
\]

\paragraph{Key Finding: OCA Does Not Degrade.}
The results reveal fundamentally different instruction-tuning dynamics from the student model setting. As shown in Figure~\ref{fig:etm_combined}, when trained with sufficient samples (100--800), the tutor model exhibits increasing MA while OCA remains stable or even \emph{improves}, rising from 93\% to 98\% across all three models (Llama, Phi, Qwen). This stands in stark contrast to student model training, where misconception acquisition consistently degraded correct-solving accuracy.

This finding is itself surprising: one might expect that training on \emph{more} misconceptions would cause \emph{more} overgeneralization and greater accuracy degradation. Instead, the opposite occurs. We hypothesize that learning multiple error patterns helps models better distinguish correct from incorrect reasoning, effectively providing contrastive signal that single-misconception training lacks. The practical implication is significant: \textbf{the tutor model does not require explicit mixing of correct examples} to satisfy both properties.

\paragraph{Sample Efficiency Constraints.}
However, as shown in Figure~\ref{fig:etm_combined}, the small-sample regime (5--80 samples) proves insufficient. With only 80 samples per misconception, only one out of ten misconceptions achieves MA above 90\%. This presents a practical challenge: a typical classroom of 20--30 students cannot provide enough data for comprehensive tutor model training. Given that misconception frequencies are heavily skewed across schools~\cite{malrules}, effective real-world tutor model training would likely require aggregating data across multiple educational institutions, highlighting the value of MalAlgoLib's controlled synthetic data for studying these dynamics.

\subsection{Step-Level Supervision: Essential for Both Models}

A natural question arises: are step-by-step solution traces merely helpful, or are they \emph{essential} for misconception acquisition? To answer this, we conducted experiments where models were trained on misconception data without intermediate steps, receiving only the final (incorrect) answer.

As shown in Figure~\ref{fig:no_steps}, the results are unambiguous: \textbf{without step-level supervision, neither the tutor model nor the student model can acquire misconceptions}. Across all three base models and all misconceptions, MA remained below 30\% even with substantial training data (up to 1600 samples for the student model, 800 for the tutor model). Several misconceptions ($M_3$, $M_5$, $M_8$) achieved less than 1\% accuracy, indicating complete failure to learn the error pattern.

For the student model without steps (Figure~\ref{fig:no_steps}, right), MA stays flat at approximately 15--25\% regardless of training size. For the tutor model without steps (Figure~\ref{fig:no_steps}, left), the situation is worse: not only does MA fail to improve, but OCA also degrades as training progresses. These findings highlight a critical insight: step-level supervision enables models to \emph{localize} errors to specific algebraic operations (e.g., the distribution step in M8) rather than diffusing erroneous patterns across the entire problem-solving process. MalAlgoLib's synthetic step-by-step traces enable us to demonstrate this requirement in a controlled setting. However, this finding poses a significant challenge for real-world applications: 
obtaining step-by-step student reasoning data  at scale requires privacy-preserving data collection infrastructure, such as secure data enclaves~\cite{safeinsights}
to capture the rich process data that misconception-aware educational AI demands.

%% file: figures/misc_acc.tex
\begin{figure*}[t!]
\centering
\includegraphics[width=0.78\linewidth]{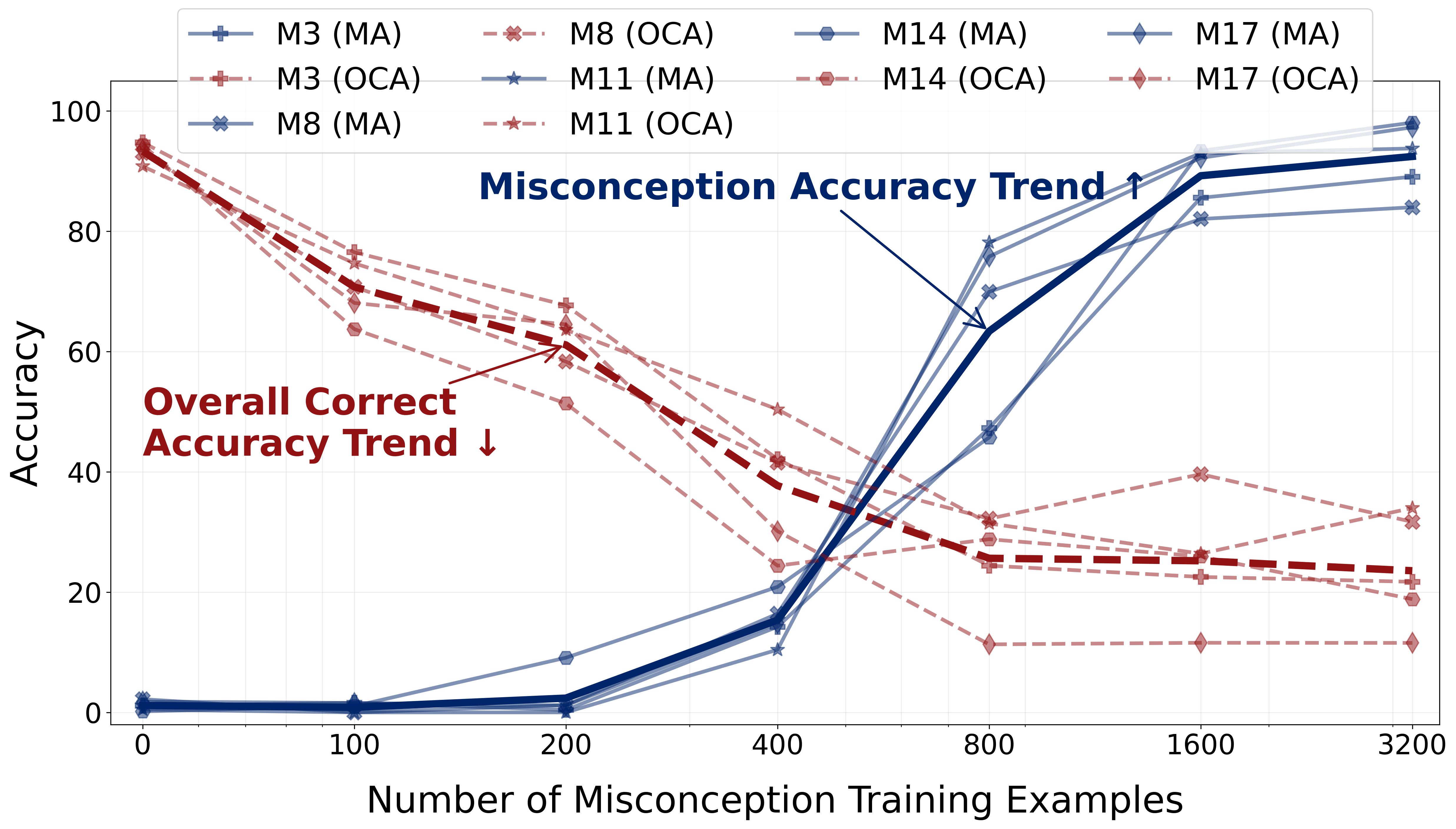} 
\caption{Novice Student Misconception Model (student model) instruction-tuning dynamics for Llama3.1-8B-Instruct \cite{llama3} across five misconceptions reveal a clear trade-off. As models acquire misconceptions (MA, solid navy), their correct-solving ability degrades (OCA, dashed burgundy). Thick lines show means; thin lines show individual misconceptions. This inverse relationship motivates investigating whether degradation occurs uniformly or differs between applicable and non-applicable problem types. Similar pattern observed for Phi4-4B-mini \cite{phi4} and Qwen3-4B-Instruct\cite{qwen3}.}
\label{fig:misconception_vs_correct_accuracy_4plots}
\end{figure*}

%% file: figures/mix_combined.tex
\begin{figure*}[h!]
\centering
\includegraphics[width=0.5\linewidth]{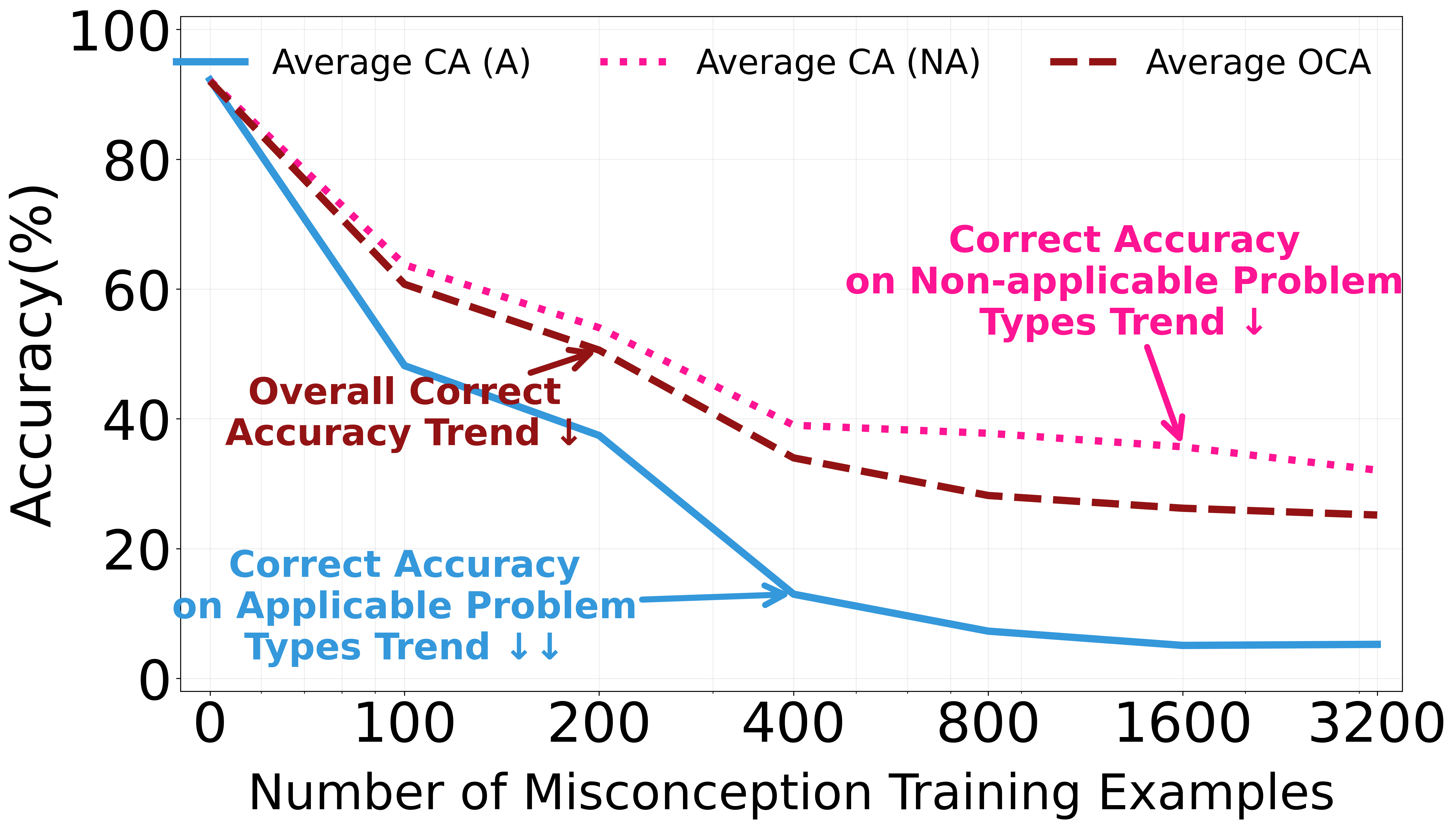} 
\hfill
\includegraphics[width=0.49\linewidth]{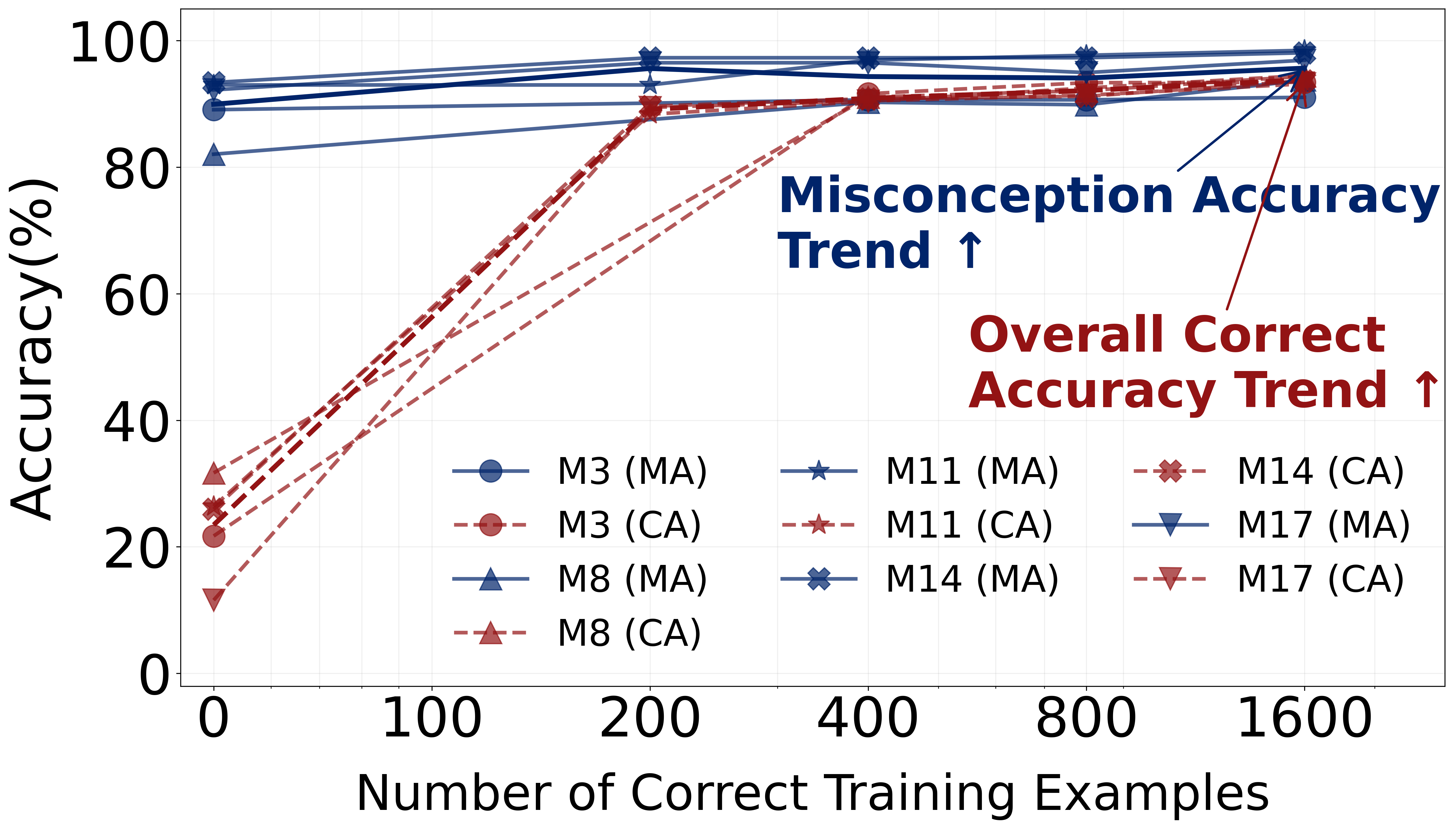} 

\caption{Left: Decomposing Novice Student Misconception Model (student model) accuracy for Llama-3.1-8B-Instruct \cite{llama3}, averaging over misconceptions, reveals that instruction-tuning degrades performance on both applicable (A, solid blue) and non-applicable (NA, dashed pink) problem types. Although $\text{Correct Accuracy (CA)}_{NA}$ declines relatively slower than $CA_A$, degrade in performance is still significant. Right: Mixing correct examples during student model instruction-tuning for Llama3.1-8B-Instruct \cite{llama3} resolves the accuracy trade-off. MA (blue) remains high (~90-95\%) demonstrating successful misconception acquisition, while CA$_{NA}$ (red) rapidly recovers with just 200-400 correct examples. Similar pattern observed for both Phi-4-4B-Mini\cite{phi4} and Qwen-3-4B-Instruct\cite{qwen3}.}
\vspace{-10pt}
\label{fig:mix_combiined}
\end{figure*}

%% file: figures/etm_combined.tex
\begin{figure*}[t]
\centering

\includegraphics[width=0.48\linewidth]{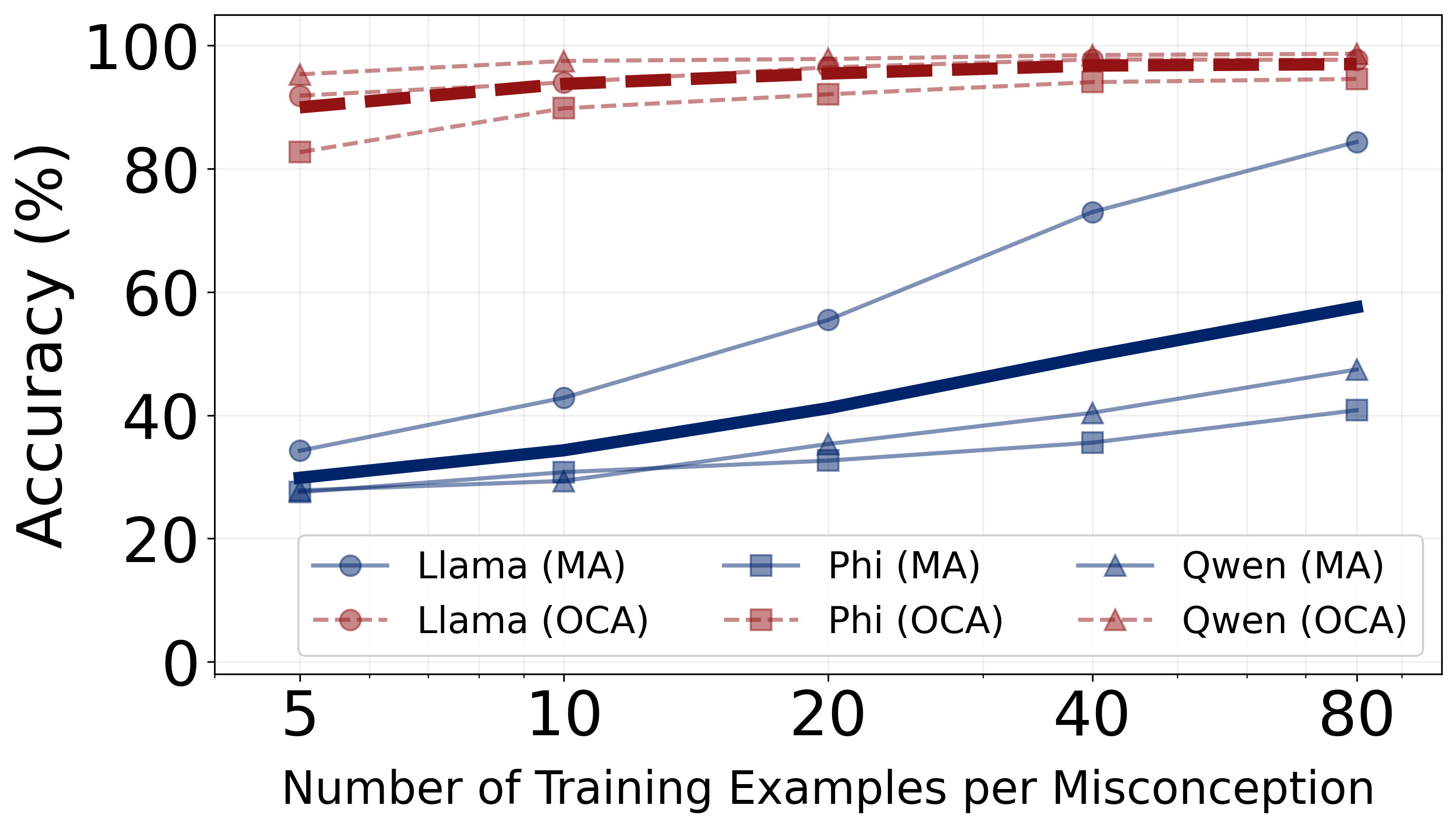}
\hfill
\includegraphics[width=0.48\linewidth]{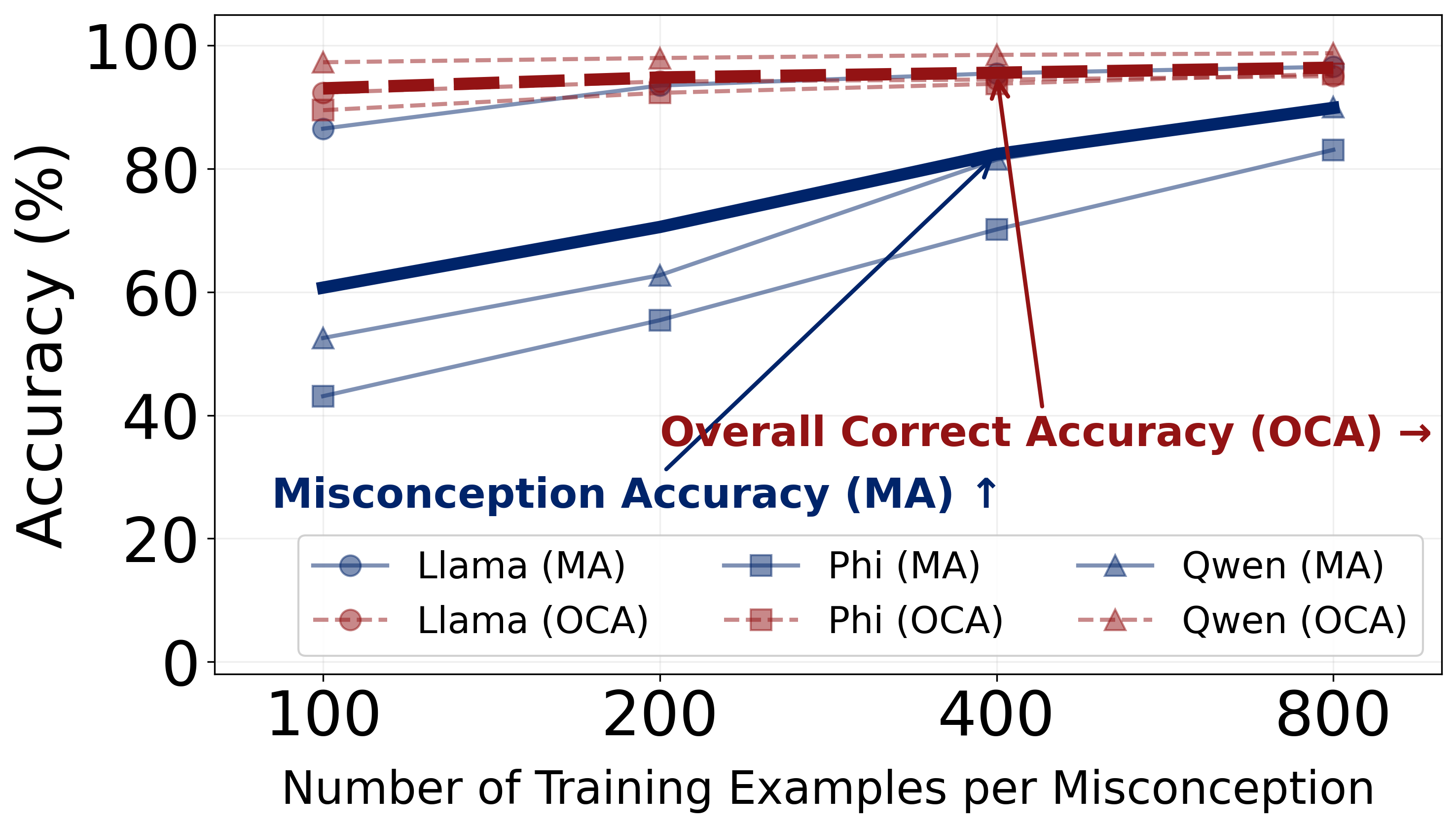}
\caption{ 
Left: Expert Tutor Misconception Model (tutor model) training on 10 misconceptions simultaneously with insufficient samples (5--80 per misconception). While OCA remains stable, average MA fails to reach the 90\% threshold for most misconceptions.  
Right: Tutor model acquisition dynamics when trained on 10 misconceptions simultaneously with sufficient samples (100--800 per misconception). Misconception Accuracy (MA, solid blue lines) increases steadily while Overall Correct Accuracy (OCA, dashed red lines) remains stable or even improves (93\% to 98\%). Results shown for Llama-3.1-8B, Phi-4, and Qwen-3 models; thick lines indicate mean trends.}
\vspace{-10pt}
\label{fig:etm_combined}
\end{figure*}

%% file: figures/no_steps.tex
\begin{figure*}[t]
\centering
\includegraphics[width=0.48\linewidth]{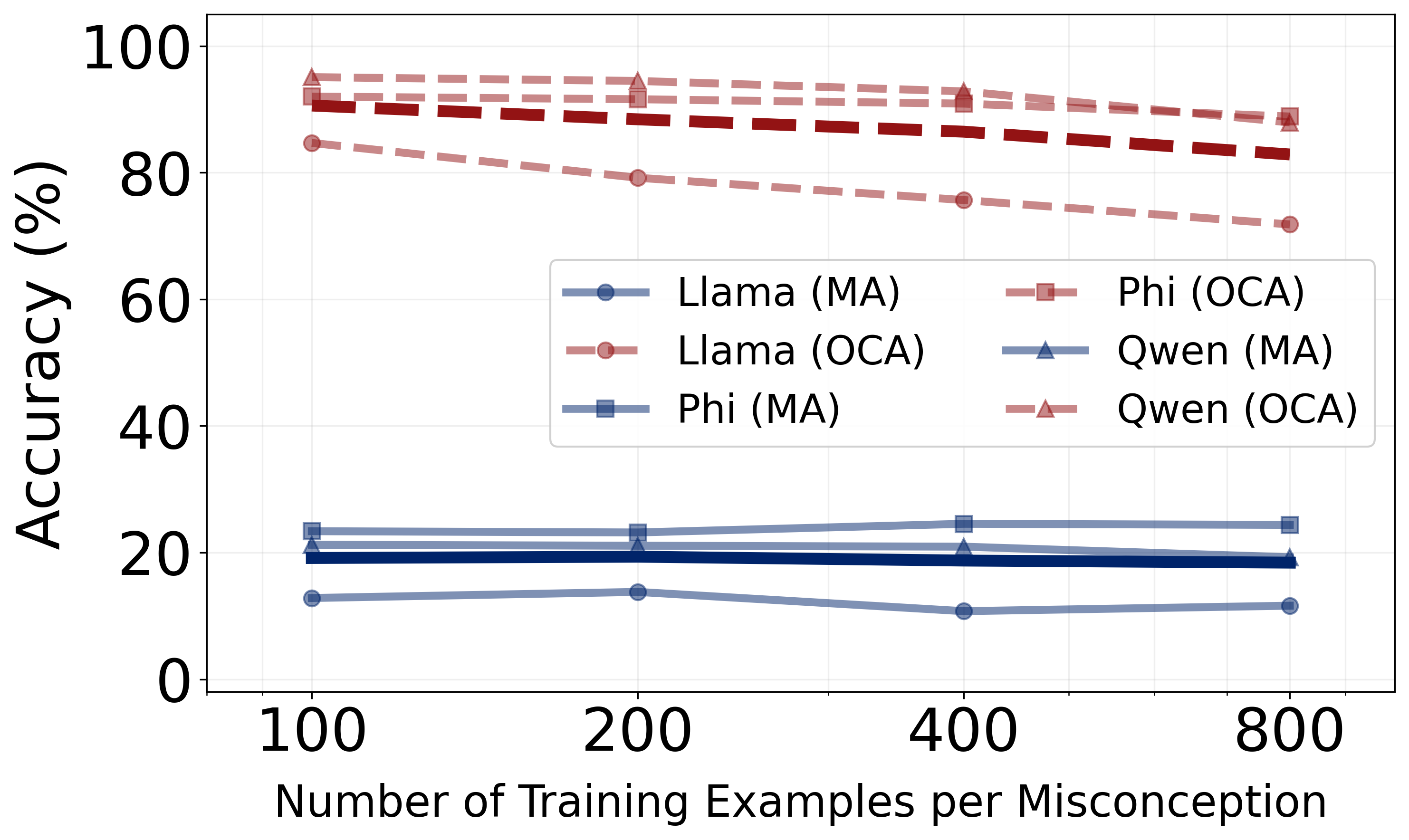}
\hfill
\includegraphics[width=0.48\linewidth]{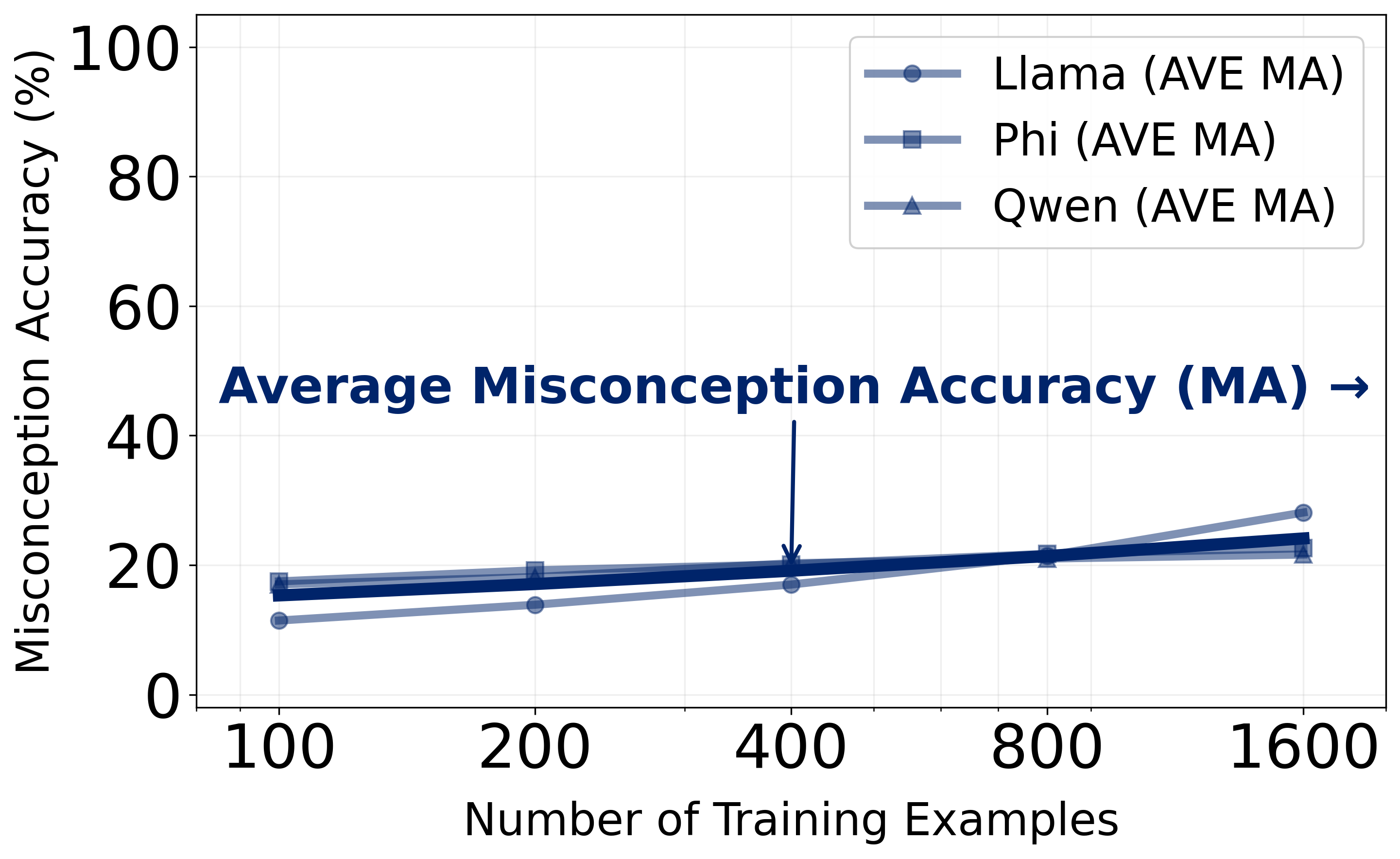}
\caption{Without step-level supervision, both the tutor model (left) and the student model (right) fail to acquire misconceptions. Even with substantial training data (up to 1600 samples for the student model, 800 for the tutor model), MA remains below 30\% across all three models (Llama, Phi, Qwen). For the tutor model without steps, OCA also degrades. This demonstrates that step-level supervision is essential for misconception acquisition.}
\vspace{-10pt}
\label{fig:no_steps}
\end{figure*}

%% file: sections/conclusion.tex
We studied the dynamics of instruction-tuning LLMs to acquire misconceptions from two complementary perspectives. The student perspective, formalized as the Novice Student Misconception Model (student model), asks whether LLMs can simulate individual students holding specific misconceptions. The tutor perspective, formalized as the Expert Tutor Misconception Model (tutor model), asks whether LLMs can capture the full spectrum of misconceptions encountered across a student population. Our experiments reveal that these two models exhibit fundamentally different instruction-tuning dynamics. For the student model, misconception acquisition comes at a cost: as models learn to replicate misconceptions, their ability to solve problems correctly degrades because the learned error overgeneralizes beyond its intended contexts. This trade-off can be mitigated by mixing correct examples into the training data. For the tutor model, no such trade-off exists; correct-solving accuracy remains stable. However, the tutor model requires substantially more training data per misconception, suggesting that effective real-world training would require aggregating data across multiple educational institutions.
Our most striking finding concerns step-level supervision. Without intermediate solution steps, neither the student model nor the tutor model can acquire misconceptions regardless of training data size. 
Together, these results, enabled by MalAlgoLib, provide an interpretable account of misconception acquisition under instruction tuning and practical guidance for training LLMs to model misconceptions while preserving correct reasoning.